\documentclass[a4paper,superscriptaddress,twocolumn,showkeys]{revtex4}

\usepackage{amsmath,amssymb}
\usepackage{subfigure}
\usepackage{framed}
\usepackage{ragged2e}
\usepackage{bm}
\usepackage{enumerate}
\usepackage{multirow}
\usepackage{graphicx}
\usepackage{url}
\usepackage{textcomp}
\usepackage{balance}
\usepackage{color}

\newcommand{\abs}[1]{\left|#1\right|} % absolute value

\newcommand{\ie}{{\it i.e.~}} % I prefer traditional way of using \it command
\newcommand{\eg}{{\it e.g.~}}

\newcommand{\etal}{{\it et al.~}}
\newcommand{\vs}{{\it vs.~}}

\newcommand{\Slabel}[1]{\label{sec:#1}} % no idea why \slabel cannot be used

\newcommand{\Sref}[1]{Sec.~\ref{sec:#1}}

\newcommand{\alabel}[1]{\label{sec:#1}} % for appendix
\newcommand{\Aref}[1]{Appendix~\ref{sec:#1}}

\newcommand{\elabel}[1]{\label{eq:#1}}

\newcommand{\Eref}[1]{Eq.~(\ref{eq:#1})}

\newcommand{\flabel}[1]{\label{fig:#1}}

\newcommand{\Fref}[1]{Fig.~\ref{fig:#1}}

\usepackage[
  pass,% keep layout unchanged 
]{geometry}

\begin{document}

\title{\Large A Dynamical Model of Twitter Activity Profiles}

\author{Hoai Nguyen Huynh}
\email{huynhhn@ihpc.a-star.edu.sg}
\homepage{https://sites.google.com/site/nelive/}
\affiliation{Institute of High Performance Computing,
Agency for Science Technology and Research, Singapore}
%1 Fusionopolis Way, \#16-16 Connexis, Singapore 138632}
\affiliation{Complexity Institute,
Nanyang Technological University, Singapore}
%18 Nanyang Drive, \#02-245 Block 2 Innovation Centre, Singapore 637723}

\author{Erika Fille Legara}
\email{legaraeft@ihpc.a-star.edu.sg}
\homepage{http://www.erikalegara.net/}
\affiliation{Institute of High Performance Computing,
Agency for Science Technology and Research, Singapore}
%1 Fusionopolis Way, \#16-16 Connexis, Singapore 138632}

\author{Christopher Monterola}
\email{monterolac@ihpc.a-star.edu.sg}
\homepage{http://www.chrismonterola.net/}
\affiliation{Institute of High Performance Computing,
Agency for Science Technology and Research, Singapore}
%1 Fusionopolis Way, \#16-16 Connexis, Singapore 138632}

\begin{abstract}
The advent of the era of Big Data has allowed many researchers to dig into
various socio-technical systems, including social media platforms. In
particular, these systems have provided them with certain verifiable means to
look into certain aspects of human behavior. In this work, we are specifically
interested in the behavior of individuals on social media platforms---how they
handle the information they get, and how they share it. We look into Twitter
to understand the dynamics behind the users' posting activities---tweets and
retweets---zooming in on topics that peaked in popularity. Three mechanisms are
considered: endogenous stimuli, exogenous stimuli, and a mechanism that dictates
the decay of interest of the population in a topic. We propose a model involving
two parameters $\eta^\star$ and $\lambda$ describing the tweeting behaviour of
users, which allow us to reconstruct the findings of Lehmann \etal (2012) on the
temporal profiles of popular Twitter hashtags. With this model, we are able to
accurately reproduce the temporal profile of user engagements on Twitter.
Furthermore, we introduce an alternative in classifying the collective
activities on the socio-technical system based on the model.

\vspace{0.5cm}
\noindent ACM Categories and Subject Descriptors:
J.2 [\textbf{Computer Applications}]: Physical Sciences and Engineering;
J.4 [\textbf{Computer Applications}]: Social and Behavioral Sciences;
I.6 [\textbf{Computing Methodologies}]: Simulation and Modeling
\end{abstract}

\keywords{Social networks, Information diffusion, Modelling}

\maketitle

\section{Introduction}
The study of information diffusion from gossip spreading
\cite{Lind_2007a,Lind_2007}, to the propagation of viral memes
\cite{Ratkiewicz_2011,Shifman_2009,Hodas.Lerman:2014}, fads, and trends
\cite{Tassier_2004,Altshuler_2012,Sano.etal:2013}, and even word-of-mouth
marketing \cite{Goldenberg_2001,Legara_2008} has become increasingly interesting
especially in this era of ``Big Data.'' Current technologies and methods have
allowed researchers to look more closely into the social network fabric---the
medium at which the proliferation of various entities takes place. Questions
relating to how fast information travels or what kind of information captures
the most audience have piqued the interest of many researchers
\cite{Asur2011,Weng_2013,Louni_2014,Ikeda_2010,Chung_2014}. Various approaches
have been implemented to shed light into these. Researchers have looked into the
role of a network's degree of connectivity, modularity, and various centrality
measures, among other things \cite{Weng_2013,Louni_2014,Ikeda_2010,Chung_2014}.
Efforts have also been put in understanding the degree of social ``influence''
of entities on each other \cite{Myers:2012,Granovetter_1978,Cosley:2010}. Many
have also investigated the nature of topics that are being diffused in a social
system. 

In this work, we propose a model that aims to capture the various aspects of
these approaches---we do not only look at the network structure in isolation,
but also augment it with particulars on the nature of the information being
spread, and the individuals' tendencies to spread such information or ``inject''
new ones. Particularly, we investigate the observations described in
\cite{Lehmann.etal:2012} on the dynamical classes of collective attention in
Twitter where they defined four groups depending on the temporal features of
their popularity dynamics. We initially introduce two free parameters intrinsic
to the users' behaviours, $\lambda$ and $\eta^\star$, where $\lambda$ quantifies
the rate of decay at which a user would spread a given information and
$\eta^\star$ is the threshold an agent has that determines whether or not he/she
propagates information from the users he/she follows. The rules defined are then
implemented in an empirical Twitter network obtained from the Stanford Large
Network Dataset Collection \cite{McAuley_2012}.

This paper is structured as follows: we first describe the data and model in
\Sref{data_model}, then present the results and discussions in
\Sref{results_discussions}, and finally summarise and establish our conclusions
in \Sref{conclusions}.

\section{Data and the model}
\Slabel{data_model}

\subsection{Data}

The dataset utilised here is a set of $115$ hashtags used by Lehmann \etal in
\cite{Lehmann.etal:2012}. It contains the time series of number of tweets and
distinct users for each of the hashtags. Each time series centers around a day
on which the number of relevant tweets attain their maximum ``popularity,'' and
spans from seven days before to seven days after the day of the peak. The full
data collected in \cite{Lehmann.etal:2012} contain $130$ million Twitter
messages appearing in the period of approximately $6$ months from November 20,
2008 to May 27, 2009. We point the readers to reference \cite{Lehmann.etal:2012}
for further details on the dataset utilised here for model fitting and
verification.\footnote{This is an aggregated dataset containing the daily 
number of tweets for each hastag and was generously provided by
Bruno Gon\c{c}alves, see Acknowledgement.} 

\subsection{The model}

\subsubsection{Definitions and rules}

The model is defined on a general network $\mathcal{N}$ with $N$ nodes, each
node representing a \emph{user}. Each user $i$ has $F_i$ ``\emph{followers}'' 
and $L_i$ ``\emph{leaders}'' whom he/she follows. This leader-follower
relationship results to a directed network. It is also worth noting that
although the Twitter network structure is dynamically changing in the
real-world, here we only consider a static structure given the relatively short
time frame  we are considering, which is two weeks. Note that when a
user $i$ follows another user $j$, the follower sees all the tweets that $j$
posts; if, on the other hand, user $i$ visits the profile page of $j$, $i$ will
not only see the tweets, but also the retweets and replies that user $j$ posts.

Three mechanisms are incorporated in our model. Two of which, exogenous and
endogenous, define the manner at which information is propagated in the system
\cite{Lehmann.etal:2012, Sornette:2008}. The endogenous process involves a
re-posting of someone else's tweet (``\emph{retweet}''), thereby
propagating/diffusing the same tweet across the social network. On the other
hand, when new information is ``injected'' in the social network system, an
exogenous process is said to have taken place. In addition to these two
mechanisms, a third one is regarded as well that accounts for the decay of the
level of the activities involving a specific topic on Twitter. To encapsulate,
our model incorporates these three processes: (1) injection of new information
into the network, (2) spreading of information in the network, and (3) decay of
information after a peak.

The key features of the model proposed are quantified in two parameters
$\eta^\star$ and $\lambda$---characterising the spreading of information and the
decay of activities in the network, respectively. The parameter $\eta^\star$
quantifies the \emph{threshold of influence} of leaders on their followers,
determining whether or not a follower would take action such as retweeting
and/or replying to a tweet, consequently exposing his/her own followers to the
information. In other words, $\eta^\star$ encapsulate the level of contagion of
a piece of information in the network. On the other hand, the parameter
$\lambda$ quantifies the \emph{rate of decay of interest} of a user in the
information after a certain point in time. It could be seen that, in our model,
the build-up in activities before a topic's peak in popularity is solely
reflected by the parameter $\eta^\star$, while the decay in activities after the
peak is the interplay between the two parameters $\eta^\star$ and $\lambda$.

To make the model results comparable with the data we have at hand, we use the
scale of one day as one time unit. The rules and flowchart of implementation of
the model are described in \Fref{model_flow_chart}. The model is updated
sequentially, \ie the state of a user $i$ at time $t$ only depends on the state
of the network \emph{before} time $t$ but not at time $t$.

\begin{figure*}[t]
\centering
\includegraphics[width=0.9\textwidth]{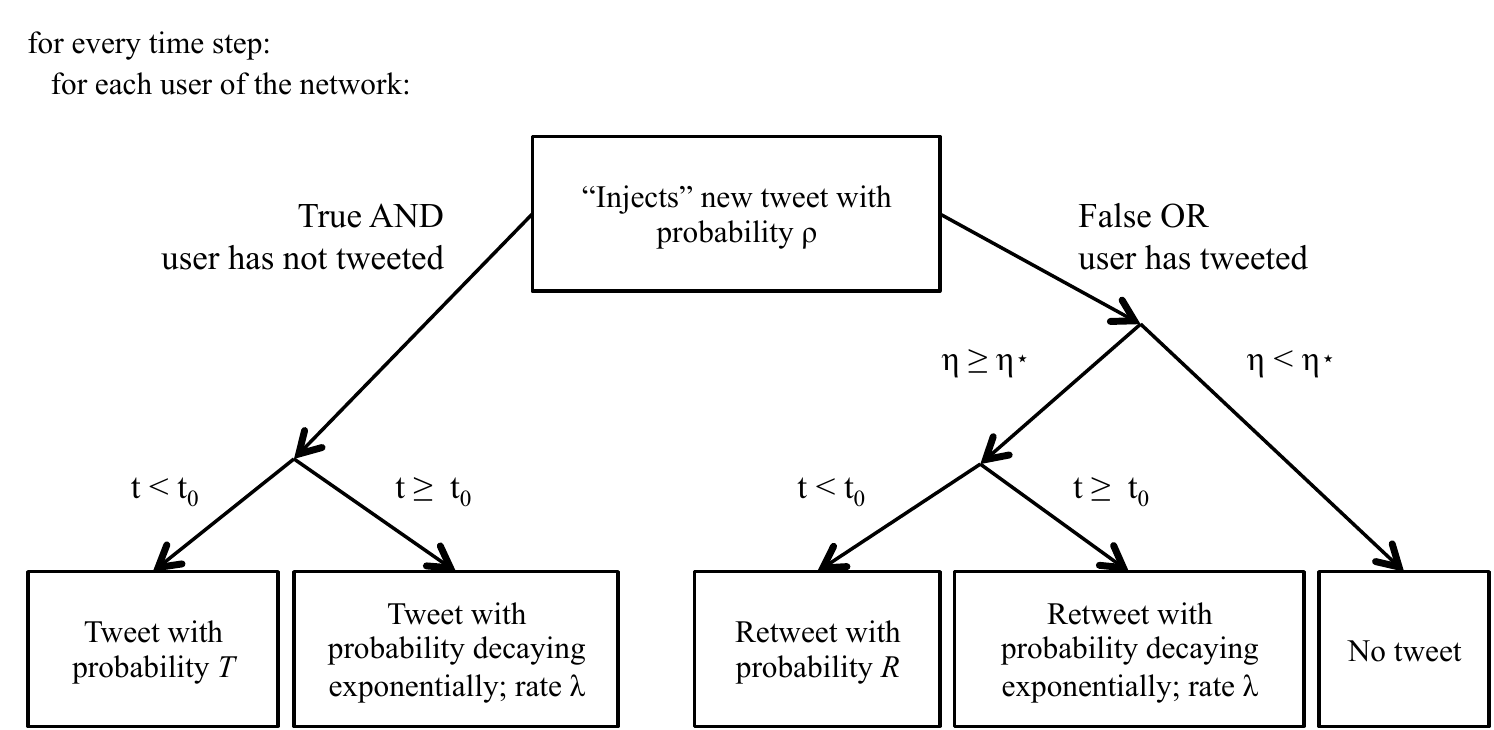}
\caption{\flabel{model_flow_chart}Rules of the model proposed in this work.
$t_0$ is the day of the peak and $\eta$ is the amount of activities by the
user's leaders accumulated after his last tweet.}
\end{figure*}

\subsubsection{Assumptions}
\Slabel{assumptions}

The model constructed makes the following assumptions on the tendency of a user
to tweet and retweet. A user posts an original\footnote{``Original'' in this
sense is used in a loose fashion. It only means that the post is not a
\emph{retweet}.} tweet if he/she is exposed to some new information outside of
his/her Twitter network, \ie from external sources (or has some original ideas
to share). A user who follows a lot of other users tends to rely solely on
his/her social network for information and, hence, retweets more often than
``injects'' new information from external sources. On the contrary, a user who
has a huge following tends to be more active in posting original ideas or new
tweets rather than just reposting others'. These assumptions on tendencies are
illustrated in \Fref{user_types}.

\begin{figure}[ht!]
\centering
\subfigure[\flabel{external_exposure}Tweeting behaviour of different types of
Twitter users based on their number of leaders and followers. Each type
corresponds to the likelihood of being exposed to external media.]
{\includegraphics[width=.45\textwidth]{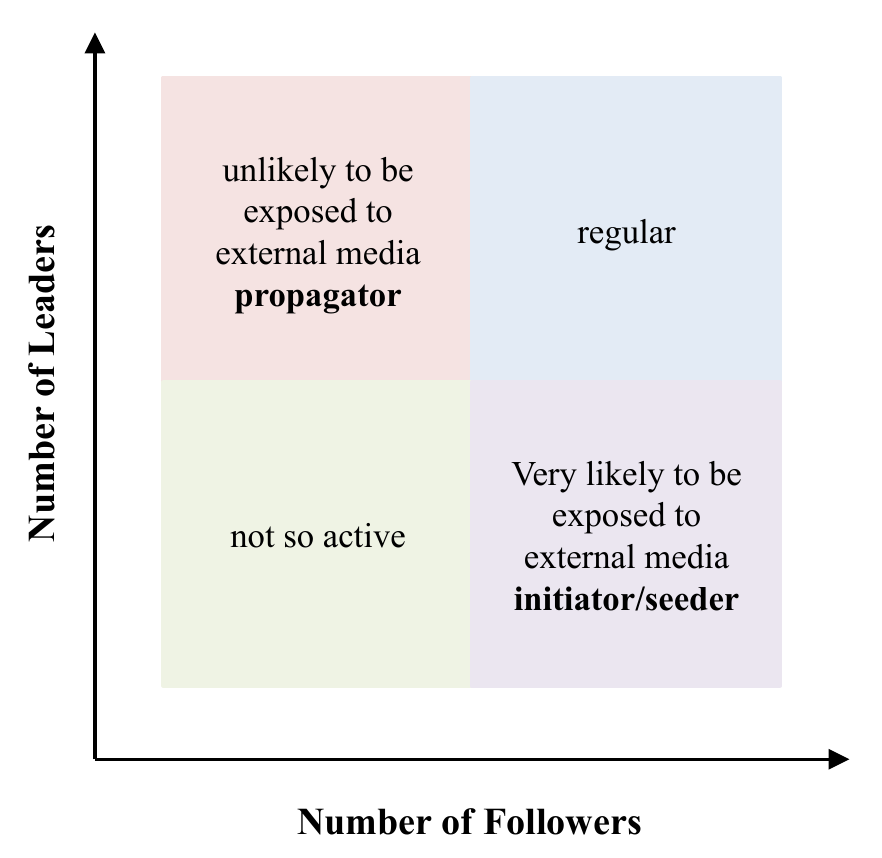}}
\subfigure[\flabel{hesitancy}Retweeting hesitancy of different types of Twitter
users based on their number of leaders and followers. The arrows indicate the
directions of increasing hesitancy, \ie when the number of leaders or followers
decreases.]{\includegraphics[width=.45\textwidth]{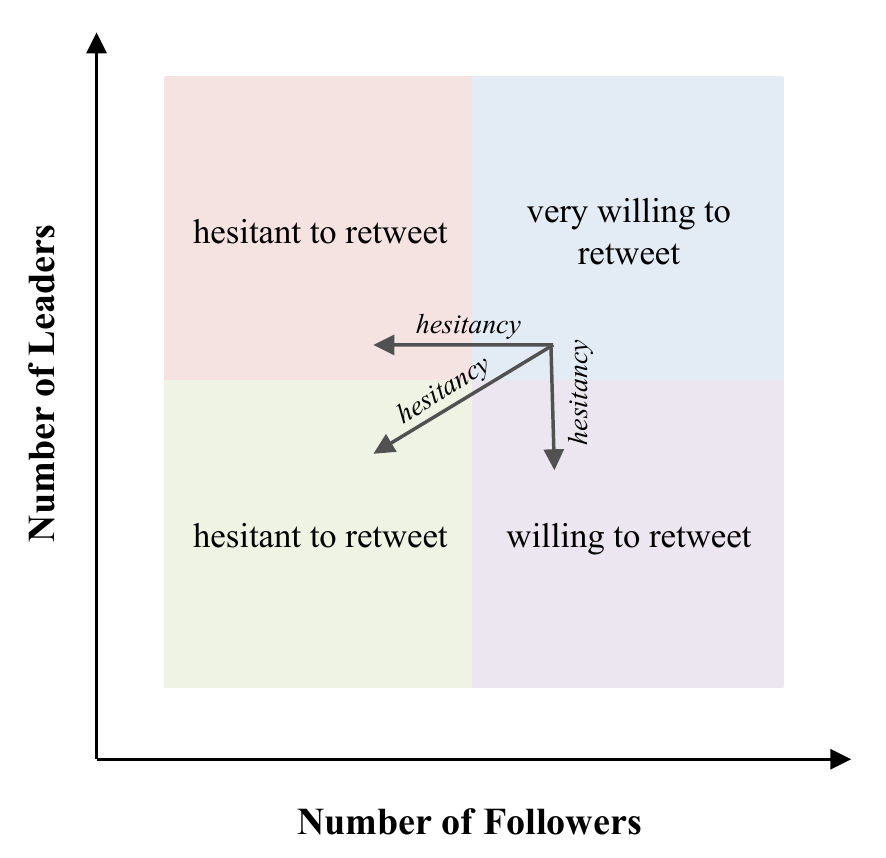}}
\caption{\flabel{user_types}Behaviour patterns of different types of users
according to their number followers and leaders.}
\end{figure}

Let us consider a user $i$ ($i=1,2,\dots,N$) who follows $L_i$ leaders $l(i,j)$
($j=1,2\dots,L_i$) and who has $F_i$ followers. The probability that user $i$ is
exposed to external sources is
\begin{equation}
\boxed{\rho_i(t) = A_i \chi(t-t_0)\text{,}}
\end{equation}
in which $A_i$ represents the activeness of $i$ in following news and
propagating to other people, and $\chi(t-t_0)$ the coverage by the media. In
general, the temporal profile of external media coverage satisfies the limiting
conditions
\begin{equation}
\left\{
\begin{aligned}
1\ge\chi(x) & \ge0~\forall x \\
\chi(0) & =1 \\
\lim_{\abs{x}\rightarrow\infty}{\chi(x)} & =0
\end{aligned}
\right.\text{.}
\end{equation}
We, however, assume that within a narrow window of time around the event, the
media coverage is consistent and stays approximately constant so that
$\chi(x\sim0)\approx1$. By the assumption described above, the activeness $A_i$
takes the form
\begin{equation}
A_i = \frac{F_i}{F_{max}} \times \left(1-\frac{L_i}{L_{max}+F_i}\right)
\end{equation}
to reflect the assumption that a user having more followers tends to be active
in following news and can introduce interesting stuff, but that is offset by
having many leaders---as in such case, the user tends to rely on the leaders for
information rather than tweeting so himself/herself as illustrated in
\Fref{external_exposure} (see, for example, \cite{Kwak.etal:2010}).

Upon external exposure, the probability of a user $i$ to tweet $T_i$ depends on:
(1) the interest of user $\sigma_i$ in the nature of the information or the
particular topic under consideration, (2) the level of interest $\tau_i(t-t_0)$
as a function of time, and (3) his \emph{hesitancy} to tweet $H_i$.

\begin{equation}
\elabel{tweeting_equation}
\boxed{T_i = \sigma_i\tau_i(t-t_0) - H_i\text{.}}
\end{equation}
The level of interest $\tau_i(t-t_0)$ is high during and before the event, and
decays with rate $\lambda$ after the event
\begin{equation}
\tau(x) = \left\{
\begin{aligned}
1 & \text{ if }x\le0 \\
\exp{(-\lambda x)} & \text{ if }x>0
\end{aligned}
\right.\text{.}
\end{equation}

The hesitancy to tweet (also retweet) depends on the number of leaders and
followers a user has as illustrated in \Fref{hesitancy}. The less leaders or
followers a user has, the more hesitant he is to retweet because of the lack of
engagement and/or motivation to do so. Hence,
\begin{equation}
\elabel{hesitancy}
H_i = \frac{1}{L_i+F_i+1}\text{.}
\end{equation}
Here, we also assume that  $\sigma=1$ indicating that we only focus on the
topics that are of interest to the users.

Next, we define the average influence of all leaders of a user $i$ as
\begin{equation}
I_i = \frac{1}{L_i} \sum_{j=1}^{L_i}{F_{l(i,j)}}\text{,}
\end{equation}
in which $F_{l(i,j)}$ is the number of followers that the leader $l(i,j)$ has.

In addition, we quantify the amount of exposure user $i$ has to the influence of
his/her leaders in the following equation:
\begin{equation}
Y_i(t) = \sum_{\substack{\text{all leaders}\\l(i,j)\text{ having}\\
\text{tweeted}\\\text{recently}\\\text{before }t}}{F_{l(i,j)}}\text{.}
\end{equation}
And the necessary condition for retweeting is
\begin{equation}
\boxed{Y_i \ge \eta^\star I_i\text{.}}
\end{equation}
Upon this condition is met, the user $i$ retweets with probability
\begin{equation}
\boxed{R_i(t-t_0) = \sigma_i\tau_i(t-t_0) - H_i\text{,}}
\end{equation}
which takes the same form as \Eref{tweeting_equation} in which $H_i$ represents
the hesitancy as described in \Eref{hesitancy}.

The number of leaders who tweeted recently, \ie after the user's last tweet and
before current time $t$, is denoted as $\eta_i(t)$. The total number of possible
retweets by user $i$ at time $t$ is given by
\begin{equation}
\boxed{\nu_i(t) =
\sqrt{\frac{\eta_i(t)}{\eta^\star}\times\frac{Y_i}{\eta^\star I_i}} \text{,}}
\end{equation}
in which we only take the integer part and take $0$ as $1$ because the number of
retweets is at least $1$ if the user retweets.

If the user retweets, it does not necessarily mean that he would retweet all $n$
tweets. The probability to retweet $R$ means that he tweets at least one tweet.
Therefore, it could be calculated that each of his $n$ possible retweets carries
probability $r=1-\sqrt[\leftroot{1}\uproot{3}n]{1-R}$.

By identifying the two key parameters $\lambda$ and $\eta^\star$, we can expect
to observe four different types of users' behaviour in response to an event, as
illustrated in \Fref{para_dist}. The four types correspond to four quadrants in
the $(\lambda,\eta^\star)$ parameter space, namely lowly contagious-slow
decaying, lowly contagious-fast decaying, highly contagious-slow decaying and
highly contagious-fast decaying.

\begin{figure}[ht!]
\centering
\includegraphics[width=0.45\textwidth]{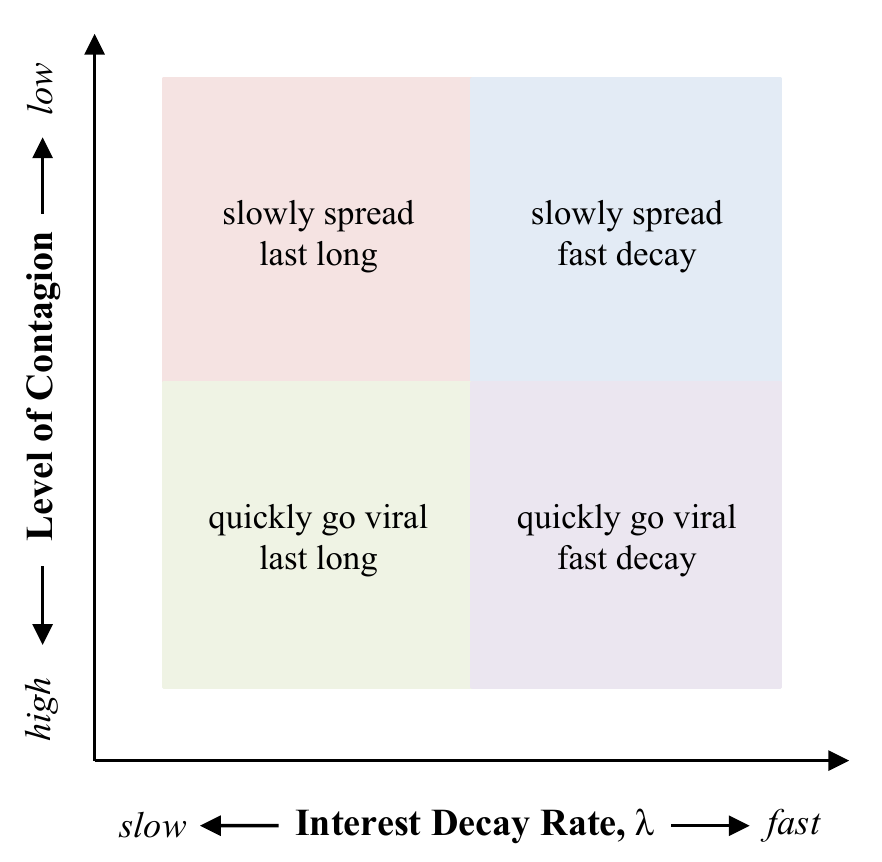}
\caption{Distribution of different types of event in the
$(\lambda,\eta^\star)$ parameter space.}\flabel{para_dist}
\end{figure}

\section{Results and discussions}
\Slabel{results_discussions}

The empirical network we use for simulation was obtained from Stanford Large
Network Dataset Collection \cite{McAuley_2012}. The entire network is a
combination of $1,000$ ego networks with $81,306$ nodes and $1,768,149$ links, a
diameter of $7$, and a clustering coefficient of $0.5653$. We run the simulation
starting from $\delta t$ days before a topic peaks in popularity $t_0$ (we also
refer to this one as ``event'') until $7$ days after $t_0$. $\delta t$ can vary
from $0$ to $7$, mimicking the fact that the amount of activities related to an
event becomes significant up to $\delta t$ days before the event. $\delta t=0$
corresponds to sudden events while a large value of $\delta t$ indicates an
anticipated one. It is noteworthy that by varying $\delta t$, we effectively
include a third parameter in our model, which characterises the injection of
information into the network.

We then scan the $(\lambda,\eta^\star)$ parameter space in the steps of
$\Delta\lambda=0.1$ ($\lambda\in[0;4]$) and $\Delta\eta^\star=1$ ($\eta^\star\in
[1;60]$) to produce different time series for the number of tweets as well as
the number of (distinct) users everyday and identify the ones that reproduce the
empirical observations by using the distance metric introduced below. Since this
is a Monte-Carlo simulation that involve generation of random numbers, we
perform $50$ runs with distinct seeds for the random number generator for each
set-up, \ie the triplet $(\delta t,\eta^\star,\lambda)$, and take the average
results.

\subsection{Validation of the model}

We compare the data generated by our model to the empirical data by calculating
the matching score of the two profiles which are quantified by the fraction of
users or tweets on a single day. In details, let $\bm P=(P_1,P_2,\dots,P_N)$ be
the profile of the tweets produced by our model, \ie $P_i$ is the fraction of
tweets on day $t_i$ within the entire period from $t_1$ to $t_N$. By definition,
we have
\begin{equation}
\sum_{i=1}^{i=N}{P_i} = 1\text{.}
\end{equation}
Similarly, $\bm Q=(Q_1,Q_2,\dots,Q_N)$ is the corresponding profile of the
tweets in the data collected by \cite{Lehmann.etal:2012}.

We compare $\bm P$ and $\bm Q$ by introducing the metric
\begin{equation}
\elabel{distance_metric}
\delta(\bm P,\bm Q) =
\frac{1}{N} \sqrt{\sum_{i=1}^{i=N}{\left(
\frac{P_i-Q_i}{\max{(P_i,Q_i)}}\right)^2}}\text{,}
\end{equation}
which quantifies the (normalised) ``distance'' between the two profiles. It is
obvious that when the two profiles are identical $\bm P\equiv\bm Q$, \ie
$P_i=Q_i~\forall i=1,2,\dots,N$, the distance is $\delta(\bm P,\bm Q)=0$. This
is a normalised measure so that the maximum possible value of $\delta$ is $1$.

In \Eref{distance_metric}, when $P_i=Q_i=0$, the term $\displaystyle
\left(\frac{P_i-Q_i}{\max{(P_i,Q_i)}}\right)^2$ does not have any contribution
to $\delta$. Finally, we set a tolerance threshold $\theta=0.04$ such that all
the terms with $P_i+Q_i\le\theta$ do not have any contribution to $\delta$.

\begin{figure*}[ht!]
\centering
\includegraphics[width=0.9\textwidth]{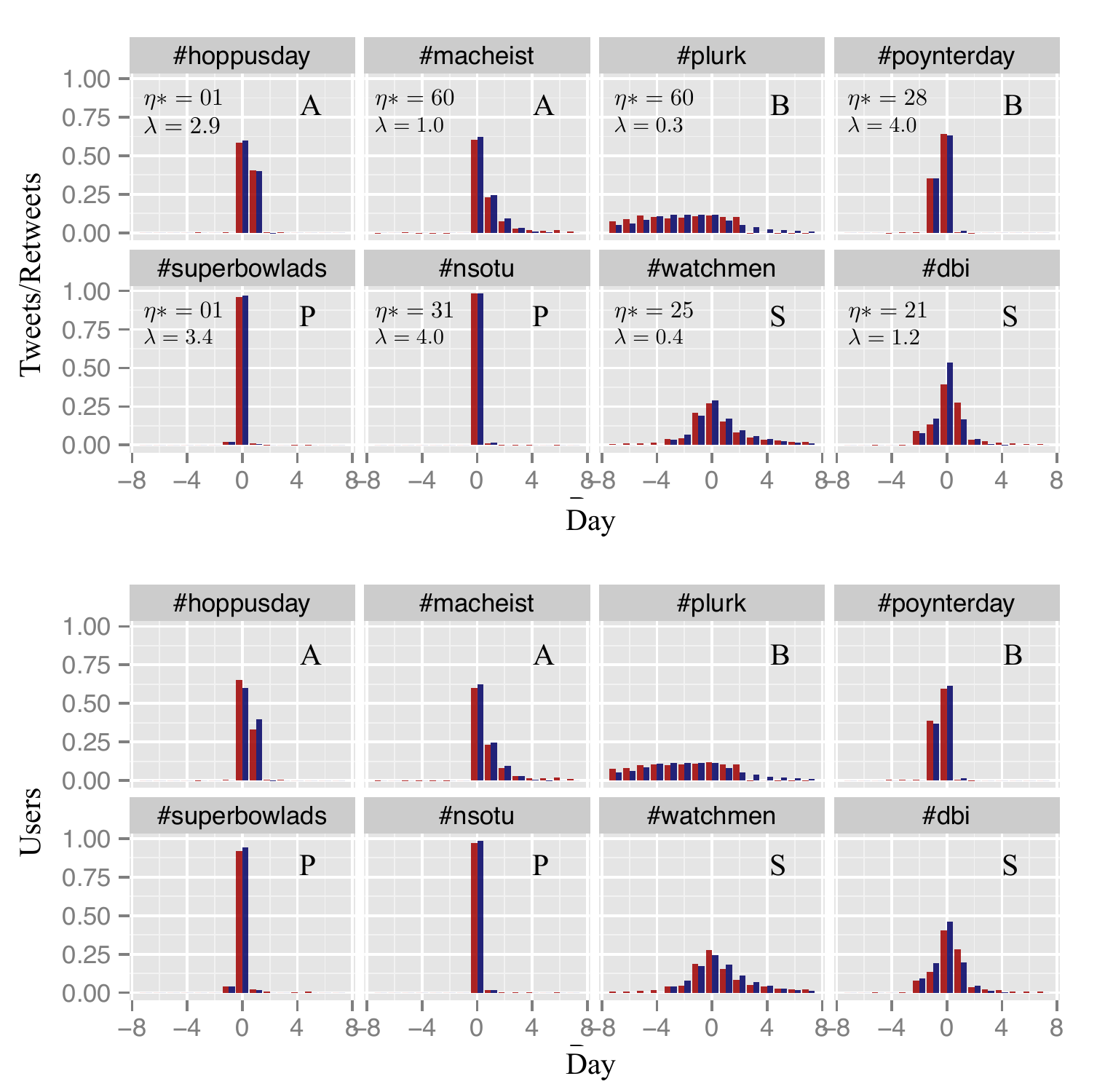}
\caption{\flabel{timeseries}Time series of activities (top) and users (bottom).
Results from the model (blue) shown together with the data (red) presented in
\cite{Lehmann.etal:2012} for classes A, B, P, and S, respectively.}
\end{figure*}

Using the metric introduced above and after visually verifying the plots
(\Fref{timeseries}), we consider measures with $\delta(\bm P,\bm Q)\le0.08$ good
and discard the rest. Of the $115$ hashtags, about $80\%$ ($88/115$) result to
good fits---both for the number of users and number of retweets. The remaining
$20\%$ fall into the groups of activities distributed before and symmetric
around the peak day \cite{Lehmann.etal:2012}, which have significant amounts of
activities distributed prior to the events. This demonstrates that the proposed
model, in spite of it being capable of capturing the main features in the
collective attention build-up and decay of users before and after the event day,
requires additional framework that would quantify the ``sense of time'' of the
users---whether or not an event is approaching \cite{Alfi.etal:2007}.
This aspect will be investigated and reported elsewhere.

It is worth noting that while it is not straightforward to know how many times a
user would tweet or retweet in a day, we have shown that our assumptions in
\Sref{assumptions} for the users' activities work well in estimating both the
number of users and retweets in most cases. Moreover, the fact that we could
reproduce the temporal profiles of activities (see \Fref{timeseries}) using our
model with only two user-intrinsic parameters and an effective third parameter
for external factors, justifies and validates our assumptions and hypotheses in
identifying the key mechanisms of information spreading in social networks.

\subsection{Classification of hashtag types}

With the estimated parameter values, we generate the plot for the distribution
of the hashtags on the two-dimensional parameter space of $\eta^\star$ and
$\lambda$, as shown in \Fref{para_cluster}. From the plot, we can observe the
clustering pattern corresponding to different types of event shown in
\Fref{para_dist}, with only a few outliers. It is quite evident that there is a
clustering of large points at the bottom left corner of the plot, which
correspond to the events that quickly go viral and last long. Those events
appear many days before the peak and generate significant amount of activities
afterward. The other three clusters contain small points signifying the events
start not so long before their peak of activities.

\begin{figure*}[t]
\centering
\includegraphics[width=0.9\textwidth]{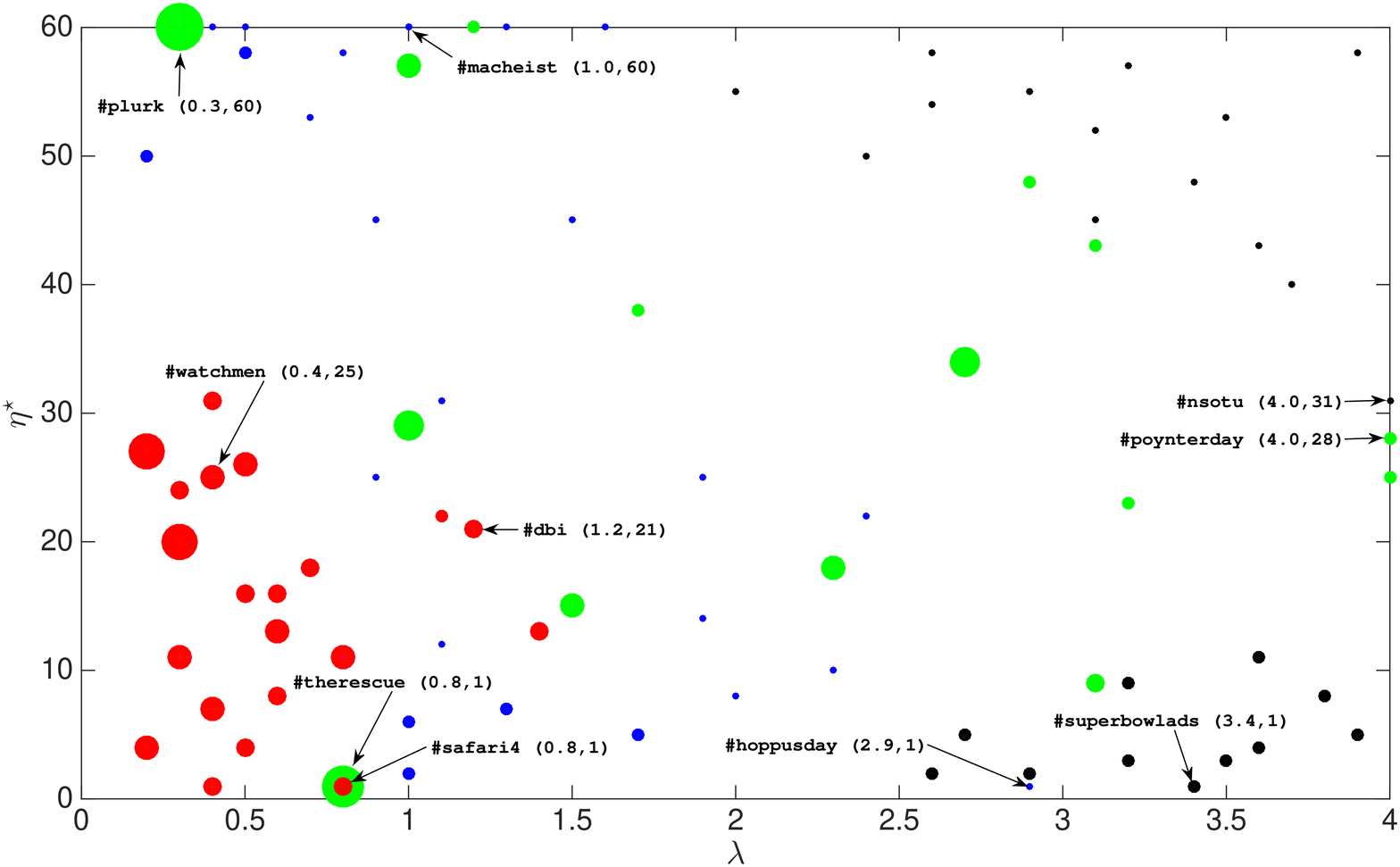}
\caption{\flabel{para_cluster}Fitted parameters $\eta^\star$ and $\lambda$
showing clustering patterns. The circles of larger size correspond to large
value of $\delta t$. The colours (online) of the data points are determined by
the classed identified in \cite{Lehmann.etal:2012}, red for S, black for P, blue
for A and green for B.}
\end{figure*}

As illustrated by the colors of the data points in \Fref{para_cluster}, we can
also observe that the distribution of the points correspond very well to the
classification of dynamical classes reported in \cite{Lehmann.etal:2012}, \ie
the points for each of the four classes can be segregated into distinct clusters
(with exception of a few points in class of activities concentrating before the
peak, see below). The four classes are called A, B, P and S, respectively, in
this work for convenience of the discussion. Class A describes events where the
associated activities are concentrated after a topic peaks in popularity. Class
B, on the other hand, refers to the events where the activities occur before the
peaks. Class P consists of events where the activities are concentrated on a
single day. Finally, Class S contains events that have significant activites
before, on and after the peak day. Our results show that the clusters described
above also reveal the existence of subclasses within each of the classes. In
\Fref{para_cluster}, we can generally identify $7$ clusters of data points (or
hashtags) which show very good correspondence to the classification in
\cite{Lehmann.etal:2012}.

From the fittings, we can observe two subgroups in the class with activities
concentrating after the peak, \ie class A (after). One group shows long range
behaviours in which the activities span over a long period of time reflected by
slow decay of interest (small $\lambda$) but high spreading threshold (large
$\eta^\star$). The other group shows short range behaviours in which the
activities span over a very short period of time reflected by low spreading
threshold (small $\eta^\star$) but very fast decay of interest (large
$\lambda$).

For the class with activities concentrating before the peak, \ie class B
(before), we also observe two subgroups. One group shows long range behaviours
in which the activities span over a long period of time reflected by long
appearance before the peak but high spreading threshold (large $\eta^\star$).
The other group shows short range behaviours in which the activities span over a
very short period of time reflected by very short appearance before the peak but
very low spreading threshold (small $\eta^\star$). 

For the class with activities concentrating at the peak, \ie class P (peak), the
values of the parameters suggest two subgroups, both of which have very fast
decay of interest (large $\lambda$). One group shows contagious behaviours in
which the events appear very shortly before the peak but generate a lot of
activities due to low spreading threshold (small $\eta^\star$). The other group
shows inert behaviours due to very high spreading threshold (large
$\eta^\star$).

The class with activities distributed symmetrically around the peak, \ie class S
(symmetric), generally has low spreading threshold (small $\eta^\star$) and slow
decay of interest (small $\lambda$).

In \Fref{timeseries}, we show the different profiles for each of the classes
described above.

\subsection{Content analysis}

After revealing the existence of the classes and subclasses of the hashtags, we
turn to looking at content of each hashtag and learn how it is related to the
apparent classification. In \Aref{hashtag_content}, we have a table showing the
hashtags together with their corresponding type and class (and subclass,
according to our results above). The table is organised in such a way that the
top rows contain the ``simple'' hashtag types, in the sense that the hashtags of
those types generally belong to one class identified by our model. The rows
further down at the bottom of the table contain more complicated hashtag types
whose tweets fall into different classes.

From the table, it could be seen that hashtags in the categories of activism
(\texttt{\#ie6}, \texttt{\#pman}) or technology (\texttt{\#safari},
\texttt{\#safari4}, \texttt{\#skype}) indicate events that capture attention in
a long period of time and make impact that keep people discussing. These events
are called for attention on a particular matter, \eg campaign or of great
interest and impact to many people, \eg technology products. The peak in these
events are usually associated with a symbolised or iconic activities on that
day, \eg rally of people in a place or release of a product. The hashtags in the
category of charity (\texttt{\#twestival}, \texttt{\#protest}) indicate events
that generate activities before a peak but soon decay after that. This is
because these events usually call for people's support to achieve a certain goal
(\eg fund raising, signature collection). And once the goal has been achieved,
people are no longer interested in the follow-up. The hashtags in the category
of marketing generally exhibit sudden appearance. That could be explained by the
strategies of marketers releasing incentives to advertise their products. But
our results show that it also depends on the type of product and how it is
advertised to determine the dynamical behaviours of people's attention to it.

The hashtags in other categories generally spread across different classes with
no easy way of relating the content to the class. Nevertheless, content type
like the Twitter (word) games spontaneously started by some user(s), which
appear in all of the classes and subclasses identified in the work, could
provide a very useful set-up to study what type of content would become popular
in a social setting \cite{Castillo.etal:2014,Rudat.Buder:2015}. Further analysis
of the meaning of the hashtags and the content of the tweet messages containing
the hashtags will be explored and reported elsewhere.

\subsection{Discussions}
 
The classification of hashtags allows us to identify their general features in
terms of how people react to the information they receive and also possibly
infer their content. Overall, class S (symmetric) occupies the bottom left
quadrant of the parameter space $(\lambda,\eta^\star)$. In this quadrant, the
threshold $\eta^\star$ is low and the rate of decay $\lambda$ is also low. They
correspond to events that can easily spread (due to low threshold) and can last
after a topic peaks in popularity (low rate of decay), \eg movie
(\texttt{\#watchmen}), technology release (\texttt{\#safari}, \texttt{\#skype})
or activism (\texttt{\#pman}). Our model in this study can reconstruct the data
very well up to $\delta=4$ days before the peak but generally falls through
beyond that. This suggests a different pattern in people's behaviour when
spreading the information when the ``sense of time'' is relevant, \ie before and
near the event associated with the information.

On the other hand, class P (peak) occupies the right half of the parameter
space, which corresponds to events that decay very quickly after the peak. They
can further be categorised into two groups: the upper one (high threshold
$\eta^\star$) corresponds to events that capture immediate attention but decay
immediately, \eg unexpected and unpopular political events
(\texttt{\#spectrial}, \texttt{\#nsotu}) or occasional media events
(\texttt{\#grammys}, \texttt{\#oscars}); and the lower one (low threshold
$\eta^\star$) corresponds to the events that spread very quickly (it appears one
or two days before the peak) and also decay very quickly, {\eg sport events
(\texttt{\#nfl}, \texttt{\#superbowl}). The remaining two classes A (after) and
B (before) can both be divided into two groups: (1) low threshold, high decay
rate; (2) high threshold, low decay rate. The difference between them is the
time the users become aware of the events. Events in class A are sudden and
people continue to discuss them due to either low decay rate (long last),
\eg lobbying marketing campaign (\texttt{\#macheist}), or low threshold (easy to
spread), \eg honouring popular stars (\texttt{\#hoppusday}). Events in class B
depict anticipation where people already discuss the topics even before their
popularities peak---this contributes to large amounts of activities before the
peak, \eg new feature of Twitter (\texttt{\#plurk}) or anticipated show
(\texttt{\#poynterday}). The events in this class, however, display scattered
pattern and in some rare cases make overlap with class S (\texttt{\#therescue}).

It needs to be emphasised that the model proposed is straightforward and
concise---carrying the heuristic and intuitive assumptions on the online
behaviours of users, given the knowledge of their social network's structure.
Yet, the model produces the dynamical behaviours observed in real data and
allows us to gain insights on the clustering of topics---telling us about the
different natures of the contents being circulated in the social media, and how
these clusters relate to the classes presented in \cite{Lehmann.etal:2012}. This
signifies that the three mechanisms included in the model are essential and
sufficient in accurately describing the dynamics behind the collection attention
of users on a Twitter network.

Knowing the relevant factors that influence the dynamics behind information
spreading and trend setting is crucial for various aspects of society which can
range from governance to politics, and marketing. Everyday, we are overwhelmed
with terabytes of information originating from various social media sources as
people share news, comments, opinions, and updates in their blogs, microblogs,
and homepages; and on Facebook, Twitter, and Instagram, among others. The key
for the stakeholders is to know how to manipulate and strategize, if possible,
their messages and campaigns such that theirs will stand out to attract
attention and not get lost in the vast sea of online information.

What we have presented herewith so far is a model that recaptures the previous
trends for certain issues and topics by describing certain attributes of the
agents involved in the social network. The next important question is whether or
not we can use this knowledge to reshape the trend profiles of the different
information types. Our work hints on the importance of knowing the kind of
audience on which a product, an idea, or a campaign has possible influence. That
aspect to some extent is quantified in our model as the parameters $\lambda$ and
$\eta*$.

\section{Conclusions}
\Slabel{conclusions}

In this work, we proposed a model using three mechanisms that underlie the
tweeting and retweeting behaviours of users on Twitter. These behaviours
correspond to perceiving and propagating information in a social network.
Despite the simplicity of the model, we are able to capture the general patterns
of behaviours observed in real data. In particular, we have not only illustrated
the four dynamical classes reported by Lehman \etal \cite{Lehmann.etal:2012} but
also demonstrated the existence of further subclasses in three of the classes.

\section{Acknowledgments}
We would like to acknowledge Bruno Gon\c{c}alves and Yang Bo for meaningful and
useful discussions. We thank Bruno for sharing with us the aggregated dataset
for use in this study. HNH thanks Chew Lock Yue at the NTU Complexity Institute
for his support. This research supported by Singapore A*STAR SERC ``Complex
Systems'' Research Programme grant 1224504056.

\bibliographystyle{abbrv}
\bibliography{references}
\balance

\onecolumngrid
\appendix
\section{Hashtag type \vs its class}
\alabel{hashtag_content}

The $88$ hashtags used in this study. They belong to $13$ types of event. Full
description of the meaning of the hashtags could be found in
\cite{Lehmann.etal:2012}.

\centering
\begin{tabular}{|p{0.1\columnwidth}|p{0.1\columnwidth}|p{0.1\columnwidth}|
p{0.1\columnwidth}|p{0.1\columnwidth}|p{0.1\columnwidth}|
p{0.1\columnwidth}|p{0.1\columnwidth}|}
\hline\hline
Class $\longrightarrow$ & \multicolumn{2}{c|}{\normalsize\bf A} &
\multicolumn{2}{c|}{\normalsize\bf B} & \multicolumn{2}{c|}{\normalsize\bf P} &
\multicolumn{1}{c|}{\normalsize\bf S} \\
\hline
Hashtag type $\downarrow$ & High $\eta^\star$ & Low $\eta^\star$ & High
$\eta^\star$ & Low $\eta^\star$ & High $\eta^\star$ & Low $\eta^\star$ & Low
$\eta^\star$ \\
\hline
\multirow{2}{*}{\parbox{0.1\columnwidth}{Activism ($2$)}} & & & & & & &
\texttt{\#ie6} \\
& & & & & & & \texttt{\#pman} \\
\hline
\multirow{3}{*}{\parbox{0.1\columnwidth}{Technology ($3$)}} & & & & & & &
\texttt{\#safari} \\
& & & & & & & \texttt{\#safari4} \\
& & & & & & & \texttt{\#skype} \\
\hline
\multirow{2}{*}{Charity ($2$)} & & & & \texttt{\#twestival} & & & \\
& & & & \texttt{\#protest} & & & \\
\hline
\multirow{5}{*}{Sport ($6$)} & & & & \texttt{\#masters} & &
\texttt{\#superbowlads} & \\
& & & & & & \texttt{\#nfl} & \\
& & & & & & \texttt{\#superads09} & \\
& & & & & & \texttt{\#nfldraft} & \\
& & & & & & \texttt{\#superbowl} & \\
\hline
\multirow{2}{*}{Honour ($3$)} & & \texttt{\#hoppusday} & & \texttt{\#poynterday}
& & & \\
& & & & \texttt{\#asot400} & & & \\
\hline
Holiday ($3$) & & \texttt{\#aprilfool s} & \texttt{\#easter} & & & &
\texttt{\#happy09} \\
\hline
\multirow{7}{*}{\parbox{0.1\columnwidth}{Convention ($10$)}} & &
\texttt{\#rp09} & \texttt{\#macworld} &
& & & \texttt{\#w2e} \\
& & \texttt{\#mix09} & & & & & \texttt{\#ces} \\
& & \texttt{\#leweb} & & & & & \texttt{\#ces09} \\
& & & & & & & \texttt{\#drupalcon} \\
& & & & & & & \texttt{\#cebit} \\
& & & & & & & \texttt{\#25c3} \\
\hline
\multirow{2}{*}{\parbox{0.1\columnwidth}{Awareness ($4$)}} & &
\texttt{\#earthday} & & \texttt{\#earthhour} & & \texttt{\#horadoplaneta} & \\
& & & & \texttt{\#therescue} & & & \\
\hline
\multirow{3}{*}{\parbox{0.1\columnwidth}{Marketing ($5$)}} & \texttt{\#glmagic}
& \texttt{\#skittles} & &
& \texttt{\#evernote} & & \\
& \texttt{\#free} & & & & & & \\
& \texttt{\#macheist} & & & & & & \\
\hline
\multirow{3}{*}{Media ($9$)} & \texttt{\#bsg} & \texttt{\#americanidol} & & &
\texttt{\#grammys} & & \texttt{\#watchmen} \\
& \texttt{\#bachelor} & \texttt{\#starwarsday} & & & \texttt{\#oscars} & & \\
& & \texttt{\#phish} & & & \texttt{\#oscar} & & \\
\hline
\multirow{4}{*}{\parbox{0.1\columnwidth}{Political ($10$)}} & & \texttt{\#g20}
& & \texttt{\#rncchair} & \texttt{\#spectrial} & \texttt{\#budget} &
\texttt{\#inaug09} \\
& & & & \texttt{\#teaparty} & \texttt{\#nsotu} & & \texttt{\#davos} \\
& & & & & & & \texttt{\#coalition} \\
& & & & & & & \texttt{\#hadopi} \\
\hline
\multirow{4}{*}{\parbox{0.1\columnwidth}{Disruption ($14$)}} &
\texttt{\#amazonfa il} &
\texttt{\#googmayh arm} & & & \texttt{\#gfail} & \texttt{\#snowmage ddon} &
\texttt{\#h1n1} \\
& \texttt{\#peace} & \texttt{\#winneden} & & & \texttt{\#gmail} &
\texttt{\#mikeyy} & \texttt{\#influenza} \\
& \texttt{\#swineflu} & & & & \texttt{\#schiphol} & & \\
& \texttt{\#bushfires} & & & & \texttt{\#blackout} & & \\
\hline
\multirow{4}{*}{\parbox{0.1\columnwidth}{Twitter ($17)$}} & \texttt{\#yourtag}
& \texttt{\#unfollow friday} & \texttt{\#tweepme} & \texttt{\#iloveyou} &
\texttt{\#nerdpick up} & \texttt{\#crapname s} & \texttt{\#dbi} \\
& \texttt{\#blogger} & & \texttt{\#firstfol low} & \texttt{\#myfirstj ob} &
\texttt{\#oscarwil deday} & \texttt{\#followme} & \texttt{\#politics} \\
& \texttt{\#socialmedia} & & \texttt{\#plurk} & & \texttt{\#3hotwords} & & \\
& & & & & \texttt{\#oneword} & & \\
\hline\hline
\end{tabular}

\end{document}